\documentclass[preprint,11pt]{aastex}
\newcommand{\beq}{\begin{equation}}
\newcommand{\eeq}{\end{equation}}
\newcommand{\e}{n_{{\rm e}}}
\newcommand{\bne}{\bar{n}_{\rm e}}
\newcommand{\de}{\delta_{\rm e}}
\newcommand{\x}{{\bf x}}
\newcommand{\dpos}{\Delta_{\rm Q}^{(2)}}

\newcommand{\dtq}{\Delta_{{\rm T}2}^{(1)}}

\newcommand{\dtqt}{\tilde{\Delta}_{{\rm T}2}^{(1)}}

\renewcommand{\k}{{\bf k}}
\newcommand{\p}{{\bf p}}

\newcommand{\xeb}{\bar{x}_{\rm e}}

\shorttitle{Patchy reionization and polarization}
\shortauthors{Weller}

\begin{document} 

\title{Inhomogeneous reionization and the polarization of the cosmic microwave background}
\author{Jochen Weller\altaffilmark{1}}
\affil{Department of Physics, University of California, Davis, CA 95616, U.S.A}
\email{weller@physics.ucdavis.edu}
\altaffiltext{1}{On leave of the Theoretical Physics Group, Blackett Laboratory, Imperial College, Prince Consort Road,  London SW7 2BZ,  U.K., e-mail: {\tt weller@physics.ucdavis.edu} }
\begin{abstract}
In a universe with inhomogeneous reionization, the ionized patches create a
second order signal in the cosmic microwave background polarization
anisotropy. This signal originates in the coupling of the free electron
fluctuation to the quadruple moment of the temperature anisotropy.
We examine the contribution from a simple inhomogeneous reionization
model and find that the signal from such a
process is below the detectable limits of the Planck Surveyor mission. However the signal is above the fundamental uncertainty limit from cosmic variance, so that a future detection with a high accuracy experiment on sub-arcminute scales is possible.
\end{abstract}
\keywords{cosmic microwave background -- cosmology:theory}
\section{Introduction}
The Microwave Anisotropy Probe and the Planck Surveyor
(\citeauthor{map}; \citeauthor{planck}; \citealt{bersanelli:96}) will provide a
precise measurement of the cosmic microwave background (CMB) temperature
anisotropy. While existing measurements of the
polarization of the CMB radiation only give crude upper limits to its
anisotropy \citep{wollack:93,partridge:97}, 
forthcoming experiments will be more sensitive to the power in the 
polarization \citep{staggs:99}. MAP is expected to make only a
statistical 
detection of the polarization anisotropy, while Planck will
measure the polarization power spectrum to high accuracy with an average pixel
sensitivity for the polarization fluctuation of around $\Delta T/ T =
5\times 10^{-6}$ (\citeauthor{planck}).		 
Since Planck will measure the power spectrum up to arcminute 
scales (\citeauthor{planck}), second order effects in the fluctuation of the
CMB radiation could become important. The study of second order effects is well
established for the temperature anisotropies
\citep{sunyaev:70,sunyaev:80,kaiser:84,ostriker:86,vishniac:87,efstathiou:88,dodelson:95,aghanim:96,hu:98,peebles:98,knox:98,haiman:99},
however for the polarization power spectrum these effects have not been
examined extensively in the past \citep{sunyaev:80,efstathiou:88} and are only
the subject of very recent investigations \citep{seshradi:98,hu:99}. It is
well known that the universe is ionized at least out to redshifts of
$z\approx5$ \citep{gunn:65} and any {\em realistic} model of how this
reionization might have taken place is thought to be inhomogeneous. The hot
ionized gas interacts with the CMB photons and results in features in the
anisotropy power spectrum.
In this {\em Letter} we will focus on the second order effects from
inhomogeneous reionization. 	 
The first order effect of reionization is an enhanced polarization anisotropy 
on large scales and a suppression on small scales \citep{zaldarriaga:97a}. The
second order effect due to homogeneous reionization is a Ostriker-Vishniac-type effect
for the polarization, where the fluctuation in the free electron density
follows the linear density variations of the overall matter
\citep{efstathiou:88,seshradi:98}. The second order contribution we are going to discuss in
this {\em Letter} is  
very similar to this effect, only that the source of the fluctuation in the
free electron density is 
different.
\section{Second order polarization anisotropy from inhomogeneous reionization}
The source of reionization is thought to be the UV radiation of early objects
like quasars and proto-galaxies as hosts of an early generation of stars. The
nuclear and gravitational energy of these objects is transformed into radiation
which subsequently ionizes the hydrogen in spheres which surround them
\citep{tegmark:94,rees:96,aghanim:96,haiman:97,haiman:98,loeb:97,silk:98,haiman:99}.  
One way to study the consequences of inhomogeneous reionization is
to use {\em effective}  
models which describe the distribution of ionized regions by a small number of
free parameters \citep{hu:98,knox:98,haiman:99}.
We adopt the model by \citet{hu:98}, which describes inhomogeneous
reionization as a set of uncorrelated patches of a certain fixed size $R$.
The number density of these patches grows with time and finally the
whole universe becomes reionized in a homogeneous way. More realistic models
include varying patch sizes \citep{aghanim:96} and contain correlations of the
ionized regions. It turns out that these
correlations lead to a somewhat different signal on smallest scales for the
temperature anisotropies \citep{knox:98}. However, the naive uncorrelated model
by \citet{hu:98} gives a good estimate of the effect on the
CMB \citep{haiman:99}. \\
We define the ionization fraction $x_{\rm e}$ to be the ratio of the number
density of free electrons $\e$ and the overall (free and bound) number density
of electrons $n$, i.e. $x_{\rm e} = n_{\rm e}/n$. Since we want to study the
effects of inhomogeneous reionization on the CMB to second order, we not only
need the mean background ionization fraction $\xeb$, we also have to know  
the variance of the distribution. \citet{hu:98} give the second moment to be
\beq
\left<x_{\rm e}(\eta_1,\x_1)x_{\rm e}(\eta_2,\x_2)\right> = \xeb(\eta_{\rm
min})\xeb(\eta_{\rm max})+ 
\xeb(\eta_{\rm min})\left[1-\xeb(\eta_{\rm max})\right]
{\rm e}^{-\frac{(\x_1-\x_2)^2}{2R^2}}  \, ,
\label{ecor}
\eeq
where $\eta \equiv \int dt/a$ is the conformal time, with $a$ the scale factor,
$\eta_{\rm min} \equiv {\rm min}(\eta_1,\eta_2)$ and $\eta_{\rm max} \equiv {\rm max}(\eta_1,\eta_2)$.
The correlation (\ref{ecor}) drops off exponentially if the distance between two
points is larger than the size of a patch $R$. If $\eta_1$ or $\eta_2$ is in
one of the homogeneous regimes (nearly neutral or complete ionization) the
correlation is just the product of the mean values.	  
In the following we will describe the inhomogeneity as the fluctuation of
the free electron number density
$\de(\eta,\x)=(n_{\rm e}(\eta,\x)-\bne(\eta))/\bne(\eta)$. We study a
universe with a $5\%$ baryon content, a Hubble constant of $H_0=100h\,{\rm
km/sec/Mpc}$, with $h=0.5$, 
critical matter density and no cosmological constant, but the results are easy
to generalize to an open or $\Lambda$ universe. The matter fluctuations are
taken to be adiabatic with an initial spectral index of $n=1.0$.\\
We are not including tensor perturbations and therefore the magnetic
component of the polarization fluctuation is zero \citep{zaldarriaga:97b}. It
is then sufficient to study the perturbations in the Stokes
parameter $Q$. If we assume the wave is traveling
into the  $z$ direction with $E_x$ and $E_y$ the amplitudes of the electric
field in $x$ and $y$ direction respectively, this parameter is given by ${\rm
Q}=\left<E_x^2\right>-\left<E_y^2\right>$. The first order
Boltzmann equation for the fluctuations in Q is 
\citep{bond:87,zaldarriaga:97b}   
\beq
\dot{\Delta}_{\rm Q}+\gamma_i\partial_i\Delta_{\rm Q}=n_{\rm
e}\sigma_{\rm
T}a\left(-\Delta_{\rm Q}+\frac{1}{2}\left[1-P_2(\mu)\right]\Pi\right)\, ,
\eeq
with $\gamma_i$ the direction of the photon momentum, $P_2(\mu)$ the
second Legendre polynomial, $\Pi=\Delta_{\rm T2}+\Delta_{\rm
Q2}+\Delta_{\rm Q0}$ the polarization tensor and the overdot refers to the
derivative with respect to conformal time. We expand in this equation the free
electron density to first order fluctuations $n_{\rm
e}(\eta,\x)=\bne(\eta)\left[1+\de(\eta,\x)\right]$ and collect all second
order terms which involve the free electron fluctuation $\de$. Since we are
only interested in the consequences of inhomogeneous reionization we do not
include other second order contributions. 
The integral solution of this second order contribution in Fourier space is then
\beq
\dpos(\eta_0,\k,\mu) =
\frac{3}{4}(1-\mu^2)\int_0^{\eta_0}g(\eta_0,\eta)e^{ik(\eta-\eta_0)\mu}
{\cal S}(\eta,\k)d\eta\, ,
\eeq
where $\mu=\cos\theta=\gamma_i k_i/k$ and $\eta_0$ is the conformal time
today. We have neglected in this solution the	 
couplings of the first order polarization fluctuations to $\de$, since the
first order temperature quadrupole will dominate these terms
\citep{seshradi:98}. The homogeneous background ionization 
history is encoded in the 
visibility function $g(\eta_0,\eta)=\dot{\tau}\exp\{-\tau(\eta_0)+\tau(\eta)\}$
with the 
differential optical depth $\dot{\tau}=\xeb n\sigma_{\rm T}a$, where $\sigma_{\rm T}$ is the Thomson scattering cross section.
The source $\cal S$ of the second order fluctuation is the mode
coupling term between the fluctuation in the free electron density and
the {\em first} order quadrupole fluctuation of the temperature,
\beq
{\cal
S}(\eta,\k)=\frac{1}{(2\pi)^{3/2}}\int\de(\eta,\k-\p)\dtq(\eta,\p)\,d^3p
\approx \frac{1}{(2\pi)^{3/2}}\de(\eta,\k)\int\dtq(\eta,\p)\,d^3p\, . 
\label{source}
\eeq
The approximation in (\ref{source}) needs explanation and we follow the
argument of \citet{seshradi:98}. The free streaming solution of the first order
quadrupole of the temperature anisotropy  is proportional to the second
spherical Bessel function, $\dtq(\eta,\p)\propto j_2(p[\eta-\eta_{\rm rec}])$,
with $\eta_{\rm rec}$ the time of recombination \citep{hu:95}. The second
spherical Bessel function can be approximated by a Gaussian with 
a peak around $p\approx p_0=3.345/(\eta-\eta_{\rm rec})$ and since we are only
interested in reionization times below $z=100$ we get $p_0<(25
0 h^{-1} {\rm Mpc})^{-1}$. The typical size of a reionized patch is of the
order $10 h^{-1} {\rm Mpc}$, which defines the scale where the free electron
fluctuation $\de$ varies the most. Therefore $\de$ is nearly
constant where the quadrupole has the largest contribution to
the integral in (\ref{source}), i.e.~$\de(\eta,\k-\p)\approx\de(\eta,\k-\p_0)$
in the relevant integration range. Furthermore the dominant $k$-range is of
the order $(10 h^{-1} {\rm Mpc})^{-1}$ and much larger than $p_0$, therefore we
can write $\de(\eta,\k-\p_0) \approx \de(\eta,\k)$.\\ 
To calculate the power spectrum we expand the fluctuation
in Q into spin-$2$ spherical harmonics and calculate the correlator of
the expansion coefficients \citep{zaldarriaga:97b}. The Stokes parameter $Q$ is
not invariant under rotation and therefore dependent on the coordinate system
we choose. However we can construct an invariant quantity
by applying a spin raising operator on Q \citep{zaldarriaga:97b}. In our
case the resulting quantity is just the electrical field $E$-type component of the polarization fluctuation.
To calculate the two-point function of the expansion coefficients we need to
know the correlator of the source $\cal{S}(\eta,\k)$, which is 
\beq
\left< {\cal S}(\eta_1,\k_1) {\cal S}(\eta_2,\k_2) \right>  
\approx 4\pi \left<\de(\eta_1,\k_1)\de(\eta_2,\k_2)\right> 
\int\limits_0^{\infty} dp\,p^2P_i(p)\dtqt(\eta_1,p)
\dtqt(\eta_2,p) \, ,
\label{sourcecol}
\eeq
where we have applied the approximation from equation (\ref{source}). Further we exploit the
fact that the first order temperature quadrupole anisotropy is {\em
uncorrelated} to the free electron fluctuation from inhomogeneous
reionization. This is not necessarily true in a universe where the patches are
correlated but the expression can be worked out as long as the underlying
fluctuations are Gaussian. The quadrupole fluctuation $\dtq(\eta,\p)$ in
equation (\ref{source}) is a random variable with amplitude and phase
depending on the initial perturbations $\psi_i(\p)$ with
$\dtq(\eta,\p)=\psi_i(\p) \dtqt(\eta,p)$ and
$\left<\psi_i(\p_1)\psi_i(\p_2)\right>=P_i(p)\delta^{(3)}(\p_1-\p_2)$ \citep{ma:95}.
The correlator in the free electron fluctuation is
given by (\ref{ecor}) and the integral in
(\ref{sourcecol}) is the correlator    
of the first order quadrupole fluctuation at unequal times. The factor
$P_i(p)$ is the COBE normalized initial power spectrum in the metric fluctuations and is given
in inflationary, adiabatic models by a power law $P_i(p)\propto p^{n-4}$, with
$n$ the spectral index. The second order anisotropy power spectrum for the
E-mode polarization is then given by 
\beq
\begin{array}{lcl}
C_{{\rm E},l}^{(2)} & = &
\frac{9}{2}(2\pi)^{9/2}R^3\frac{(l+2)!}{(l-2)!}
\displaystyle \int dk\,d\eta_1\,d\eta_2\, g(\eta_0,\eta_1)g(\eta_0,\eta_2)Q_{\rm
P}(\eta_1,\eta_2)I(\eta_1,\eta_2)\times \\
& &
\displaystyle
k^2\exp\left[-\frac{k^2R^2}{2}\right]\frac{j_l(x_1)}{x_1^2}\frac{j_l(x_2)}{x_2^2}\, ,
\end{array}
\label{cel}
\eeq
with $I(\eta_1,\eta_2)=\xeb^{-1}(\eta_{\rm min})-1$, $j_l(x)$ the spherical
Bessel functions, $x_i=k(\eta_0-\eta_i)$ and the correlation in the quadrupole
at unequal times $Q_{\rm P}(\eta_1,\eta_2) \equiv \int
dp p^2 P_i(p) \dtqt(\eta_1,p) \dtqt(\eta_2,p)$, i.e. $Q_{\rm
P}(\eta_0,\eta_0) \propto C_{T,2}$. We should notice that the limits of the
time integrals are the start and end of the reionization process,
i.e. we integrate over the time during which the inhomogeneities appear. 
For the background 
reionization history we assume that the mean ionization fraction $\xeb$ grows
from zero to unity between the redshifts $z^*-\delta z^*/2$ and $z^*+\delta
z^*/2$, which we will choose appropriately.

\section{Results}
We have calculated (\ref{cel}) with a modified version of CMBFAST
\citep{seljak:96}. The COBE normalized polarization power spectrum is shown in
figs.\ref{fig:1} and \ref{fig:2} for different parameters. In fig.\ref{fig:1}
the background reionization is given by $z^*=50$ and $\delta z^*=20$. The
dotted line refers to the {\em first} order contribution. We recognize the
feature on large scales which appears because of the homogeneous reionization
background. 
\placefigure{plot1}
The thick long-dashed, solid and short-dashed lines refer to the
second order contribution due to inhomogeneous reionization. The long-dashed
line is a model with a patch size of $R=20 h^{-1} {\rm Mpc}$, the solid line
for $R=10 h^{-1} {\rm Mpc}$ and the short-dashed line for $R=5 h^{-1} {\rm
Mpc}$.\\ 
To understand the behavior of the second order effect we will perform two
approximations to the integral in (\ref{cel}). First we realize that the
expression $g(\eta_0,\eta_1)g(\eta_0,\eta_2)I(\eta_1,\eta_2)$, for
reasonable parameters of the reionization history, is a $2$-dimensional
function with a narrow peak. This allows us to perform the
two time integrations in (\ref{cel}) and calculate the integrand at a
certain time $\eta^{*}$ multiplied by an area factor $(\delta\eta)^2$.
The second feature we exploit is that
the spherical Bessel function $j_l(x)$ has a
tight peak at $l=x$ for large multipole moments $l$, so we can approximate it with a $\delta$-function.
Therefore we get
\beq
C_{{\rm E},l}^{(2)} \approx
\frac{9}{2}(2\pi)^{9/2}\frac{(l+2)!}{(l-2)!}
\displaystyle \Theta_0^3 l^{-2} Q_{\rm
P}(\eta^*,\eta^*)I(\eta^*,\eta^*)g^2(\eta_0,\eta^*)(\delta\eta)^2{\rm
e}^{-\frac{l^2\Theta_0^2}{2}} j_l(l)\, ,
\eeq
with $\Theta_0=R/(\eta_0-\eta^*)$ as the angular size of a patch as seen
today.   
The only unknown quantity in this expression is the effective time period
$\delta\eta$. For the case with $R=10 h^{-1} {\rm Mpc}$ we get a good fit with $\delta\eta=6.5
h^{-1} {\rm Mpc}$, which is in the range of the width of the narrow peaked
function $g(\eta_0,\eta_1)g(\eta_0,\eta_2)I(\eta_1,\eta_2)$. We recovered a
similar behaviour for the whole range of reasonable reionization parameters. Although we can not predict the amplitude of the second
order power spectrum in (\ref{cel}) analytically, the shape is well
approximated by $l(l+1)C_{{\rm E},l}^{(2)} \propto l^4 {\rm
e}^{-l^2\Theta_0^2/2}j_l(l)$. This describes essentially an $l^4j_l(l)$ rise with
some cut-off around the scale $l \approx \sqrt{2}/\Theta_0$ and we see in
fig.\ref{fig:1} how the peak of the signal moves to larger multipoles $l$
when we decrease the patch size $R$. We also recognize
that the amplitude is proportional to $\Theta_0^3$ and therefore the larger 
the patch size $R$ the larger the 
power in the second order anisotropy from inhomogeneous reionization.\\
In fig.\ref{fig:2} we 
have plotted the results for a model with a more realistic background
reionization history \citep{haiman:99}. 
\placefigure{plot2}
In this case reionization takes place around
$z^*=10$ with a time period of $\delta z^*=5$. The long-dashed line
corresponds to a patch size of $R=10 h^{-1} {\rm Mpc}$, the solid line to $R=5
h^{-1} {\rm Mpc}$ and the short dashed-line to $R=1h^{-1} {\rm Mpc}$.
One clearly sees that the power in the anisotropy for this late
time reionization is much smaller than in the case for earlier times in
fig.\ref{fig:1}. This is because the visibility $g(\eta_0,\eta^*)$ is much smaller for the short and late time reionization phase. Again the the peak moves to the right when the patch size is
decreasing. For inhomogeneous reionization with {\em correlated} patches we
expect the same behavior as for the temperature anisotropy power spectrum
as discovered by \citet{knox:98}. In these models the power in the anisotropies
is much wider distributed over the multipole moments $l$ than for an uncorrelated scenario, but the magnitude
is very similar.

\section{Conclusion}
\begin{figure}
\label{plot1}
\plotone{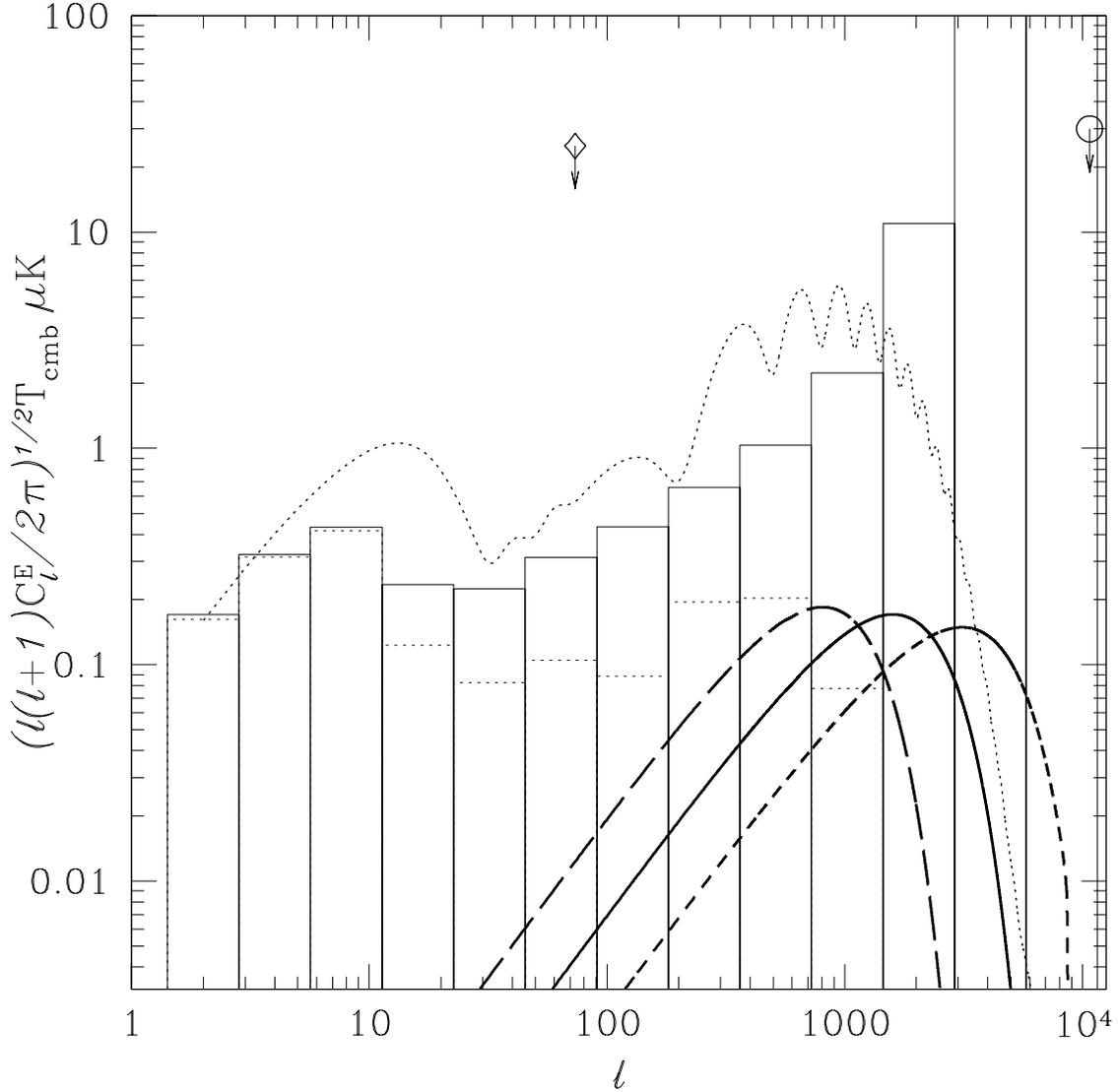}
\caption{The CMB polarization anisotropy power spectrum for a
model with an effective reionization time $z^*=50$ and $\delta z^*
= 20$. The dotted line is the first order contribution. The second order
signal from inhomogeneous reionization is the long-dashed line
for a patch size of $R=20 h^{-1} {\rm Mpc}$, the solid line for $R=10 h^{-1}
{\rm Mpc}$ and the short-dashed line for $R=5 h^{-1} {\rm Mpc}$. The diamond
and circled data points are the $95\%$ confidence upper limit from the
Saskatoon anisotropy experiment and the VLA $8.4$ GHz CBR project,
respectively. The histogram shows the logarithmically binned 
uncertainty limit from the polarization measurement with the
Planck Surveyor. The horizontal dotted lines in the histogram correspond to
the uncertainty levels due to cosmic variance only.\label{fig:1}} 
\end{figure}

\begin{figure}
\label{plot2}
\plotone{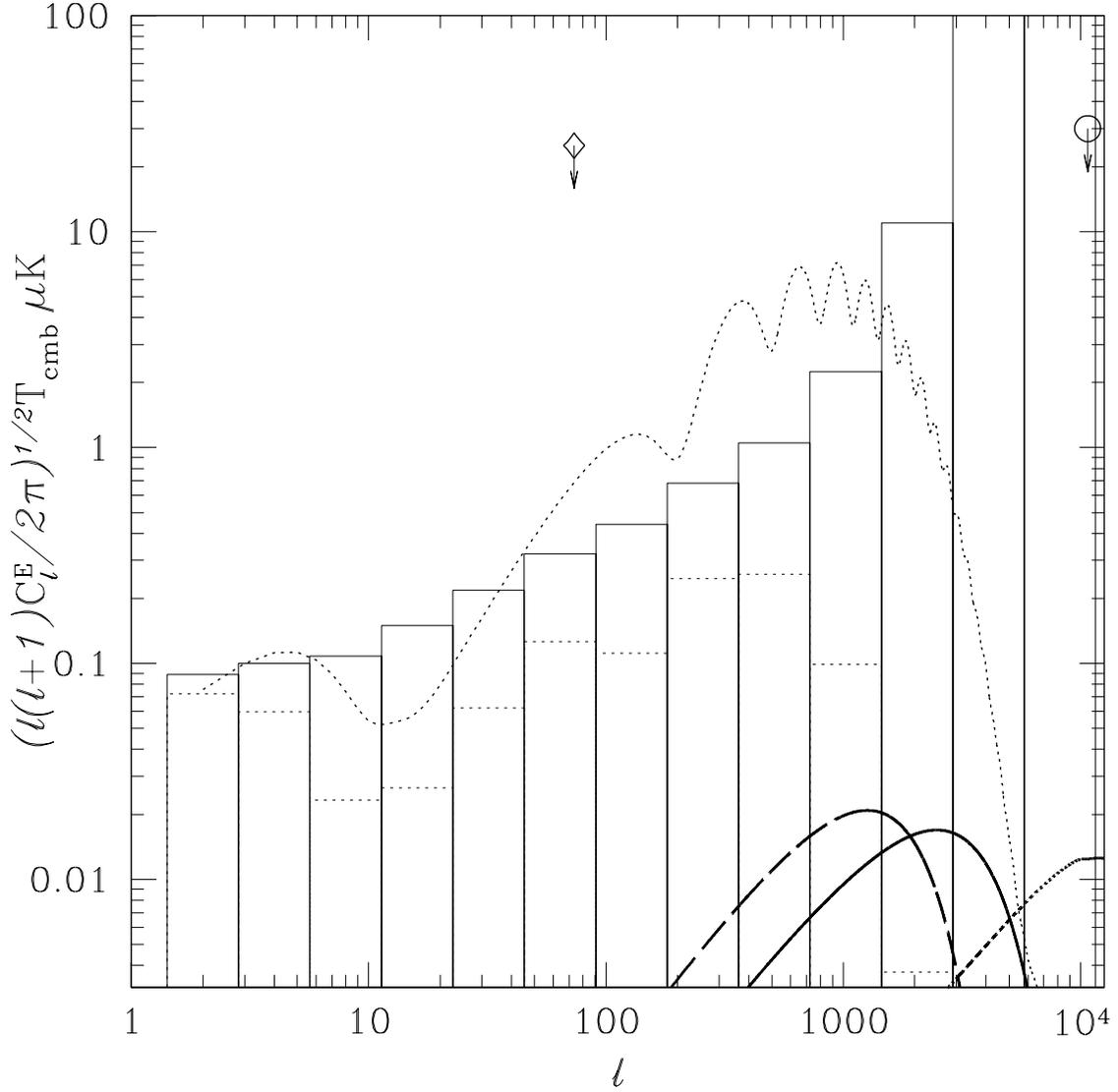}
\caption{The same plot as figure \ref{fig:1} but for an effective
reionization at $z^*=10$ and $\delta z^*=5$. The patch sizes are $R=10 h^{-1}
{\rm Mpc}$ for the long-dashed line, $R=5 h^{-1} {\rm Mpc}$ for the solid line
and  $R=1h^{-1} {\rm Mpc}$ for the short-dashed line.\label{fig:2}}
\end{figure}

In figs.\ref{fig:1} and \ref{fig:2} we find that the second order signal
dominates over the first order signal only for very large multipoles. Even for 
an unrealistic reionization model like in fig.\ref{fig:1} the second order
contribution is relevant only on scales smaller than $5$ arcsec. One might
hope, that once the cosmological parameters are estimated by the first
order temperature power spectra and other experiments, one can reveal the
nature of the second order effects.  The Ostriker-Vishniac effect for polarization is
of the same order or smaller than the signal from inhomogeneous reionization, 
dependent on the ionization parameters \citep{seshradi:98}. However this
effect is completely determined by the linear power spectrum and therefore can
be removed like the first order contribution.\\ 
At the present
polarization has not been detected in the CMB and the 
measurements give only crude upper limits 
\citep{staggs:99}. We have given the $95\%$ confidence upper limits from the
Saskatoon anisotropy experiment \citep{wollack:93} and the VLA $8.4$ GHz CBR
project \citep{partridge:97} as a circled and a diamond point in
figs.\ref{fig:1} and \ref{fig:2}. The most accurate future experiment which
will measure the polarization anisotropy is the Planck Surveyor. Its High
Frequency Instrument (HFI) is expected to measure the polarization to high
accuracy in its $143$ and $217$ GHz channels. The average sensitivity $\Delta
T/T$ to the linear polarization per pixel and the angular resolution is
$3.7\times10^{-6}$ and $8.0$ arcmin, and $8.9\times 10^{-6}$ and
$5.5$ arcmin, respectively (\citeauthor{planck}). We have 
calculated the expected polarization signal sensitivity due to cosmic variance,
beam size and instrument noise with the methods described in
\citep{knox:95,bond:97,bond:98}. In figs.\ref{fig:1} and \ref{fig:2} we show the sensitivity histograms, resulting from a logarithmic bin by weighted average. One recognizes that the noise 
levels are all above the second order signal from inhomogeneous reionization,
so that even with a fairly wide binning strategy, the sensitivity of the Planck
Surveyor is not large enough to reveal these signals. The dashed horizontal
lines in the histograms in figs.\ref{fig:1} and \ref{fig:2} are the
logarithmically binned uncertainty contributions just from cosmic variance,
which are given by $(\Delta C_l)^2=2C_l^2/(2l+1)$, with full sky coverage. Cosmic variance is the fundamental uncertainty limit and describes the 
fact that we can only observe one universe with only $2l+1$ modes  of a
certain multipole moment $l$. It is
clear from figs.\ref{fig:1} and \ref{fig:2} that the second order contribution
from inhomogeneous reionization is above these limits for large multipoles. 
Therefore a high accuracy polarization measurement on sub-arcminute scales
could reveal such a signal. But this depends on how well one 
can remove polarization foregrounds and the magnitude of other foreground-type
second order contributions, like the Sunyaev-Zel'dovich effect
\citep{hu:99}. However, there might be the possibility to disentangle all these effects by exploiting small scale measurements of the matter distribution and the CMB anisotropies, including the information from polarization.
\acknowledgments
The author gratefully acknowledges helpful conversations with N.~Aghanim,
A.~Albrecht, L.~Knox, A.~Lewin, J.~Magueijo and M.~Zaldarriaga. Further I
thank U.~Seljak and M.~Zaldarriaga for the use of CMBFAST. 
JW is supported by the German Academic Exchange Service (DAAD).

\def\jnl#1#2#3#4#5#6{{#1 #2, {#4\/} {#5}, #6 }}
\def\jnltwo#1#2#3#4#5#6#7#8{\hang{#1, {#4\/} {#5}, #6; {\it ibid} {\bf #7} #8 (#2)}}
\def\prep#1#2#3#4{{#1, #2, preprint (#4)}} 
\def\proc#1#2#3#4#5#6{{#1 #2, in {\it #4\/}, #5, eds.\ (#6)}}
\def\book#1#2#3#4{\hang{#1, {\it #3\/} (#4, #2)}}
\def\jnlerr#1#2#3#4#5#6#7#8{\hang{#1 [#2], {\it #4\/} {#5}, #6.
{Erratum:} {\it #4\/} {\bf #7}, #8}}
\def\prl{Phys.\ Rev.\ Lett.}
\def\pr{Phys.\ Rev.}
\def\pl{Phys.\ Lett.}
\def\np{Nucl.\ Phys.}
\def\prp{Phys.\ Rep.}
\def\rmp{Rev.\ Mod.\ Phys.}
\def\cmp{Comm.\ Math.\ Phys.}
\def\mpl{Mod.\ Phys.\ Lett.}
\def\apj{ApJ}
\def\apjl{ApJ (Letter)}
\def\aa{{\rm A\&A}}
\def\aap{Astron.\ Ap.}
\def\cqg{Class.\ Quant.\ Grav.} 
\def\grg{Gen.\ Rel.\ Grav.}
\def\mn{MNRAS}
\def\ptp{Prog.\ Theor.\ Phys.}
\def\jetp{Sov.\ Phys.\ JETP}
\def\jetpl{JETP Lett.}
\def\jmp{J.\ Math.\ Phys.}
\def\zpc{Z.\ Phys.\ C}
\def\cupress{Cambridge University Press}
\def\pup{Princeton University Press}
\def\wss{World Scientific, Singapore}
\def\oup{Oxford University Press}

\clearpage

\begin{thebibliography}{}
\bibitem[Aghanim et al.(1996)]{aghanim:96}{\jnl{Aghanim, N., D\'esert, F.~X.,
Puget, J.~L., \& Gispert, R.}{1996}{}{\aa}{311}{1}}

\bibitem[Bersanelli et al.(1996)]{bersanelli:96}Bersanelli, M.~et al. 1996,
Phase A study for the {\em Cobras/Samba} Mission (Paris: European Space Agency D/SCI(96)3) 

\bibitem[Bond \& Efstathiou(1987)]{bond:87}{\jnl{Bond, J.~R., \& Efstathiou,
G.}{1987}{}{\mn}{226}{655}}

\bibitem[Bond, Efstathiou \& Tegmark(1997)]{bond:97}{\jnl{Bond, J.~R.,
Efstathiou, G., \& Tegmark, M.}{1997}{}{\mn}{291}{L33}}

\bibitem[Bond, Jaffe \& Knox(1998)]{bond:98}{\jnl{Bond, J.~R., Jaffe, A.~H.,
\& Knox, L.}{1998}{}{\pr}{D 57}{2117}}

\bibitem[Dodelson \& Jubas(1995)]{dodelson:95}{\jnl{Dodelson, S., \& Jubas, J.~M.}{1995}{}{\apj}{439}{503}} 

\bibitem[Efstathiou(1988)]{efstathiou:88}{\proc{Efstathiou, G.}{1988}{}{Large Scale Motions in the Universe: A Vatican Study Week}{Princeton University Press, Princeton, New Jersey}{Rubin, V.~C., \& Coyne, G.~V.}}

\bibitem[Gruzinov \& Hu(1998)]{hu:98}{\jnl{Gruzinov, A., \& Hu, W.}{1998}{}{\apj}{508}{435}}

\bibitem[Gunn \& Peterson(1965)]{gunn:65}{\jnl{Gunn, J.~E., \& Peterson,
B.~A.}{1965}{}{\apj}{142}{1633}} 

\bibitem[Haiman \& Loeb(1997)]{haiman:97}{\jnl{Haiman, Z., \& Loeb,
A.}{1997}{}{\apj}{483}{21}} 

\bibitem[Haiman \& Loeb(1998)]{haiman:98}{\jnl{Haiman, Z., \& Loeb,
A.}{1998}{}{\apj}{503}{505}} 

\bibitem[Haiman \& Knox(1999)]{haiman:99}{\proc{Haiman, Z., \& Knox, L.}{1999}{}{Microwave Foregrounds}{ASP, San
Francisco}{de Oliveira-Costa, A., \& Tegmark, M.}}

\bibitem[Hu \& Sugiyama(1995)]{hu:95}{\jnl{Hu, W., \& Sugiyama, N.}{1995}{}{\apj}{444}{489}}

\bibitem[Hu(1999)]{hu:99}{\prep{Hu, W.}{1999}{}{astro-ph/9907103}}

\bibitem[Kaiser(1984)]{kaiser:84}{\jnl{Kaiser, N.}{1984}{}{\apj}{282}{374}}

\bibitem[Knox(1995)]{knox:95}{\jnl{Knox, L.}{1995}{}{\pr}{D 52}{4307}}

\bibitem[Knox, Scoccimarro \& Dodelson(1998)]{knox:98}{\jnl{Knox, L., Scoccimarro, R., \& Dodelson, S.}{1998}{}{\prl}{81}{2004}}

\bibitem[Loeb(1997)]{loeb:97}{\prep{Loeb, A.}{1997}{}{astro-ph/9704290}}

\bibitem[Ma \& Bertschinger(1995)]{ma:95}{\jnl{Ma, C.~P., Bertschinger, E.}{1995}{}{\apj}{455}{7}}

\bibitem[MAP(1996)]{map} MAP homepage at:  {\tt http://map.gsfc.nasa.gov/}

\bibitem[Ostriker \& Vishniac(1986)]{ostriker:86}{\jnl{Ostriker, J.~P., \& Vishniac, E.~T.}{1986}{}{\apjl}{306}{L51}}

\bibitem[Partridge et al.(1997)]{partridge:97}{\jnl{Partridge, R.~B., Richards,
E.~A., Fomalont, E.~B., Kellerman, K.~I., \& Windhorst, R.~A.}{1997}{}{\apj}{483}{38}}

\bibitem[Peebles \& Juskiewicz(1998)]{peebles:98}{\jnl{Peebles, P.~J.~E., \& Juskiewicz, R.}{1998}{}{\apj}{509}{483}}

\bibitem[Planck Surveyor(1996)]{planck} Planck Surveyor home page at: {\tt
http://astro.estec.esa.nl/Planck/ }

\bibitem[Rees(1996)]{rees:96}{\prep{Rees, M.~J.}{1996}{}{astro-ph/9608196}}

\bibitem[Seljak \& Zaldarriaga(1996)]{seljak:96}{\jnl{Seljak, U., \&
Zaldarriaga, M.}{1996}{}{\apj}{469}{437}}

\bibitem[Seshradi \& Subramanian(1998)]{seshradi:98}{\jnl{Seshradi, T.~R.,
\& Subramanian, K.}{1998}{}{\pr}{D 58}{063002}}

\bibitem[Silk \& Rees(1998)]{silk:98}{\jnl{Silk, J., \& Rees,
M.~J.}{1998}{}{\aap}{331}{L1}}  

\bibitem[Staggs, Gundersen \& Church(1999)]{staggs:99}{\proc{Staggs, S.~T.,
Gundersen, J.~O., \& Church, S.~E.}{1999}{}{Microwave Foregrounds}{ASP, San
Francisco}{de Oliveira-Costa, A., \& Tegmark, M.}}

\bibitem[Sunyaev \& Zel'dovich(1970)]{sunyaev:70}{\jnl{Sunyaev, R.~A., \&
Zel'dovich, Ya.~B.}{1970}{}{Ap\&SS}{7}{3}}

\bibitem[Sunyaev \& Zel'dovich(1980)]{sunyaev:80}{\jnl{Sunyaev, R.~A., \&
Zel'dovich, Ya.~B.}{1980}{}{\mn}{190}{413}}

\bibitem[Tegmark, Silk \& Blanchard(1994)]{tegmark:94}{\jnl{Tegmark, M., Silk,
J., \& Blanchard, A.}{1994}{}{\apj}{420}{484}}

\bibitem[Vishniac(1987)]{vishniac:87}{\jnl{Vishniac, E.~T.}{1987}{}{\apj}{322}{597}}

\bibitem[Wollack et al.(1993)]{wollack:93}{\jnl{Wollack, E.~J., Jarosik,
N.~C., Netterfield, C.~B., Page, L.~A., \& Wilkinson, D.}{1993}{}{\apjl}{419}{L49}}

\bibitem[Zaldarriaga(1997)]{zaldarriaga:97a}{\jnl{Zaldarriaga,
M.}{1997}{}{\pr}{D 55}{1822}}

\bibitem[Zaldarriaga \& Seljak(1997)]{zaldarriaga:97b}{\jnl{Zaldarriaga, M., \&
Seljak, U.}{1997}{}{\pr}{D 55}{1830}}

\end{thebibliography}
\end{document}